\documentclass{elsart}
\usepackage[dvips]{graphicx}
\newcommand{\icm}{\ensuremath{\textrm{ cm}^{-1}}}

\usepackage{amssymb}

\begin{document}

\begin{frontmatter}

\title{Temperature dependence of the spectral weight in p-- and n--type cuprates: 
a study of normal state partial gaps  and  electronic kinetic energy }


\author {N.~Bontemps, R.P.S.M.~Lobo}
\address{Laboratoire de Physique du Solide (UPR5 CNRS), 10 rue Vauquelin, 75231 Paris cedex 05, France}

\author {A.F.~Santander-Syro}
\address{Laboratoire de Physique des Solides Universit\'e Paris--Sud, 91405 Orsay, France}

\author {A.~Zimmers}
\address {Center for Superconductivity Research, Department of Physics, 
University of Maryland, College Park, Maryland 20742, USA}

\begin{abstract}
The  optical conductivity of CuO$_2$ (copper-oxygen) planes in p and n--type cuprates thin films at various doping levels 
is deduced from highly accurate reflectivity data. The temperature dependence of the real part $\sigma_{1}(\omega)$ 
 of this optical conductivity and the corresponding spectral 
weight allow to track the opening of a partial gap in the normal state of n-type Pr$_{2-x}$Ce$_x$CuO$_4$ 
(PCCO), but not of p--type Bi$_2$Sr$_2$CaCu$_2$O$_{8+\delta}$ (BSCCO) cuprates. This is a clear difference between 
these two families of cuprates, which we briefly discuss. 
In BSCCO, the change of the electronic kinetic energy $E_{kin}$ -- deduced from 
the spectral weight-- at the 
superconducting transition is found to cross over from a conventional BCS behavior (increase of $E_{kin}$ below $T_c$) 
to an unconventional behavior (decrease of $E_{kin}$ below $T_c$) as the free carrier density decreases.
This behavior appears to be linked to the energy scale over which spectral weight is lost 
and goes into the superfluid condensate, hence may be related to Mott physics.

\end{abstract}

\end{frontmatter}

\section{Introduction}
\label{Introduction}

High $T_c$ superconductors (cuprates) 
deviate in many fundamental aspects from conventional superconductors. 
  One basic difference is that high $T_c$ superconductors can be regarded as 
  doped Mott insulators \cite{Imada00} due to their common structure, namely
the  CuO$_2$ planes, whether  electron-- or hole--doped. Strictly speaking, cuprates are charge transfer 
insulators when undoped, but the physics is close \cite{Zhang88}.   
More and more theoretical insight has been brought up recently on the relevance of Mott physics 
\cite{Phillips04,Millis05,Haule06}. 
In this paper, we present in a synthetic way a fairly large body of our work dealing 
with the electromagnetic response 
 of the CuO$_2$ planes (the issue of the c-axis optical conductivity is not addressed at all)  
 in the normal (N) and superconducting (SC) state of cuprates, which may be related to Mott physics.

 Cuprates exhibit 
superconductivity in a moderate doping range, and 
 an anomalous normal (above $T_c$) state characterized 
in a given doping and temperature range by a normal state gap or "pseudogap".
Here we will rather use the former expression, namely a normal state 
partial  (NSP) gap, defined  as follows: 
{\it i)} there is a 
depression of the density of states (DOS) at the Fermi level; {\it ii)} in  k-space, there are regions 
where quasi-particles are well 
defined at the Fermi surface, and others where they are not seen (gapped 
Fermi surface) \cite{note1}.

In hole--doped materials, the literature on the NSP  gap ("pseudogap") is huge
\cite{TimuskStatt99}.  Angle-resolved 
photoemission (ARPES) experiments performed in the BSCCO material show that 
this NSP gap opens along the $(0,\pi)$ direction in $k$ space \cite{Norman98} 
and evolves smoothly into the superconducting gap as the temperature $T$ is 
lowered. However,  gap-like features related to the NSP gap do not appear
in the optical conductivity: only 
 a  decrease in the optically defined scattering rate develops below typically 
 800~cm$^{-1}$ \cite{Puchkov96} and no loss of spectral weight (SW) at low energy 
 is balanced by an 
 upward shift  to high energy, although it could be expected in the case  
 of a density  gap \cite{Santander02,Millis04}.

On the electron doped side \cite{Fournier98}, the most studied material is 
Nd$_{2-x}$Ce$_x$CuO$_{4}$ (NCCO). 
Opposite to hole--doped cuprates, a NSP gap was observed to induce a low-to-high energy spectral 
weight transfer in the optical conductivity of non superconducting single crystals ($x=0$ to $0.125$).
This was taken as an evidence of the opening of a high energy ``pseudogap'' at temperatures well above 
the N\'eel temperature 
\cite{Onose04}. ARPES measurements mapping the Fermi surface at low temperature 
for $x=0.15$, show that intensity is suppressed where the nominal Fermi surface crosses the
magnetic Brillouin zone boudary \cite{Armitage02}, thus suggesting a gap.
In a previous paper \cite{Zimmers05}, we presented our data on the optical conductivity of 
Pr$_{2-x}$Ce$_{x}$CuO$_{4}$ (PCCO) films which can be made superconducting in the underdoped regime. 
Our films display a similar NSP gap as in NCCO single crystals; this gap however survives in 
superconducting samples, in contrast with \cite{Onose04} and it closes at x=0.17 Ce concentration, 
well inside the superconducting dome.
We analyzed this in terms of a quantum phase transition, consistent with transport evidence 
in similar samples \cite{Dagan04}. We suggested that this gap originates 
from a spin density wave, consistent with ARPES \cite{Armitage02}.
A question then arises: is a gap being seen in the conductivity of n-type and not in that of 
p-type an  indication that p and n-type cuprates are different, or is it merely contingent upon 
specific details of the Fermi surface? 

Superconductivity in cuprates has been addressed in various ways.
In the last years, several papers have discussed   
the change $\Delta SW_{SC-N}$ = $SW_{SC}-SW_{N}$ of the total spectral weight 
-- with obvious notations -- in 
BSCCO,  when the system goes from normal to superconducting 
\cite{Molegraaf02,Santander03,Santander04,Deutscher05}. 
One reason for this continuing interest is that it has been argued for a long time that
the spectral weight can be related to the electronic kinetic energy $E_{kin}$ 
\cite{Millis04,vdM00,Norman02}. 
Hence $\Delta SW_{SC-N} = - \Delta E_{kin,SC-N}$ . 
In cuprates, due to their larger gap and smaller Fermi energy compared to standard metals, 
 $\Delta E_{kin,SC-N}$  is not immeasurably small if estimated from 
standard BCS theory \cite{deGennes}. 
A phase diagram representing the change 
$\Delta E_{kin,SC-N}$ as a function of the doping level showed 
that $\Delta E_{kin,SC-N}$ can be either negative, thus unconventional 
(on the underdoped side of the phase diagram), or positive, thus BCS-like (on the overdoped side).  
These results have been now established by two independent groups in BSCCO 
\cite{Molegraaf02,Santander03,Santander04,vdmsurdop}. They  appeared momentarily controversial 
\cite{Boris04}.  The  data analysis, which was the issue  at stake, has now been clarified 
 \cite{Kuzmenko05,Santandercom} and there is no counter evidence so far of the phase diagram shown
 in \cite{Deutscher05}. 
This brings up important questions such as: is the sign change of $\Delta E_{kin,SC-N}$ 
 a signature of the system 
becoming closer to the Mott insulator? Is the high temperature superconductivity mechanism 
unique throughout the whole phase diagram?

The paper is organized as follows: 
we recall the samples which we have studied, the experimental conditions for the reflectivity measurements 
and the optical conductivity derived 
from these measurements. We define the spectral weight and the assumptions to be made 
 to relate the spectral weight to $E_{kin}$. We then contrast the experimental data in the n-type PCCO samples
 and the p-type BSCCO samples.
We show how
the spectral weight in the normal state is a powerful tool in order to study the low lying electronic states, 
despite the fact that the optical conductivity is an average over the Fermi surface.
We then show the systematic variation of the spectral weight change 
 in the case of the n-type BSCCO cuprate 
as the system becomes superconducting. We end up with the discussion of the relevance of  high energy states 
(Mott physics).

\section{Optical conductivity, spectral weight and kinetic energy}
\label{Optical conductivity, spectral weight and kinetic energy}

The spectral weight $W$ is the integral of the real part $\sigma_{1}(\omega)$ of the conductivity 
over a given frequency range:

\begin{equation}
	W(T)=\int_{\omega_{1}}^{\omega_{2}} \sigma_{1,xx}(\omega, T) d\omega 
\label{SW}
\end{equation}
The subscript $xx$ refers to the direction $x$ of the current flow when induced 
by an electromagnetic field along the same direction $x$. If $\omega_1=0$ 
and $\omega_2=\infty$, Eq.1 is just the usual f-sum rule, which takes into account {\it all} optical 
transitions:  $W$ is a constant independent of any parameter.
Taking into account only the free carriers' contribution to $\sigma_{1}(\omega)$ 
by using 
a Hamiltonian where all other degrees of freedom are (ideally) integrated out, 
thus restricting to a subset of optical transitions, a restricted sum rule writes \cite{Millis04}:
\begin{equation}
	SW(T)=\int_{0}^{\infty} \sigma_{1,xx}(\omega, T) d\omega = \frac{\pi e^{2} a^{2}}
						{2 \hbar^{2} V} E_{K}(T)
\label{EqSumRule}
\end{equation}

where $e$ is the electron charge, $a$ the in-plane lattice constant, $V$ the volume 
of the unit cell.  
$E_{K}$ is given by:
\begin{equation}
	E_{K}(T)=\frac{2}{a^{2} N} \sum_{k}\frac{\partial^{2}\epsilon_{k}}{\partial k_{x}^{2}}
				n_{k}(T)
\label{EqEK}
\end{equation}
where $N$ is the number of $k$ vectors, $\epsilon_{k}$  the dispersion 
from the kinetic energy part of the Hamiltonian, and $n_{k}(T)$  the 
momentum distribution function.
 
In a single band,  tight-binding model with nearest-neighbor interactions, this quantity 
represents the electronic kinetic energy $E_{kin}$:
\begin{equation}
	E_{kin}(T) = -E_{K}(T)	
\label{EqEkin}
\end{equation}

Equation 4 is exact in the case of nearest-neighbor hopping, \cite{Millis04,vdM00,Norman02} 
and still holds, within a correction factor, 
if next nearest neighbor interactions are taken into account \cite{vdM03}.
In this description, the system has one conduction band. 
There exists localized excitations such as phonons 
 whose spectral weight can be shown to be negligible compared to 
the electronic contribution. 
The full band stucture, however, allows 
 a number of interband transitions.
Therefore, from a practical point of view, rather than integrating up to infinity (Eq.2), 
an upper cut-off frequency $\omega_2$ must be selected 
in order to capture only  the 
conductivity associated to the carriers in the conduction band and avoid the contribution of 
interband transitions at high energy. 
This cut-off frequency has been discussed at length \cite{Millis05,Santander04,Santander06}.
We found
that a good test that spectral weight with a given $\omega_2$ frequency represents correctly the behavior of 
free carriers is that the final result does not depend qualitatively, if not exactly quantitatively, 
on the choice of  $\omega_2$ (within an acceptable range) \cite{Millis05,Deutscher05}. 
Another procedure to define the cut-off frequency can also be used, yielding a similar figure for 
$\omega_2$ \cite{Hwang04}.

\section{Experimental}
\label{Experimental}

The samples studied are epitaxial thin films and all the relevant experimental details are decribed elsewhere
\cite{Millis05,Santander04}.
BSCCO thin films were epitaxially grown by r.f. magnetron 
sputtering on (100) SrTiO$_3$ substrates. Their maximum critical temperature 
(defined at zero resistance) is $\sim 84$~K \cite{Zorica99}. These samples are:
 $T_c=70$~K, underdoped; $T_c=80$~K, near optimal doping; and  
$T_c=63$~K, overdoped. Their resistance are shown in fig.1, left panel. 
PCCO films were epitaxially grown by pulsed-laser deposition on SrTiO$_3$ substrates 
 \cite{Maiser98}. These samples have nominal Ce compositions 
$x=0.11$ (not superconducting  down to 4 K); $x=0.13$ ($T_c=15$~K); $x=0.15$ ($T_c=21$~K); and  
$x=0.17$ ($T_c=15$~K). Their resistivities 
are shown in fig.1, right panel. 
\begin{figure}
  \begin{center}
      \includegraphics[width=0.9\textwidth]{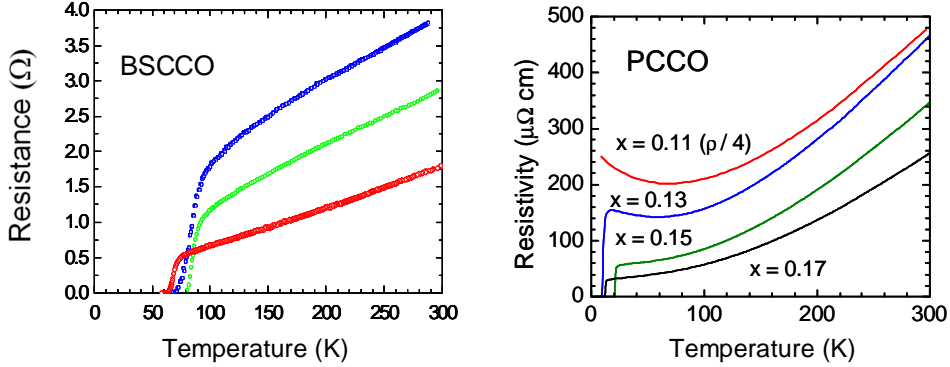}
  \end{center}
  \caption{\label{Fig1}
    Left panel: resistance of the three BSCCO films; 3 $\Omega$ correspond typically to 200 $\mu\Omega$.cm.
    The most resistive film is the underdoped one ($T_c$=70~K); the intermediate one is nearly optimally doped 
    ($T_c$=80~K); the least resistive film is overdoped ($T_c$=63~K).
     Right panel: resistivity of the four 
  PCCO films labelled in the figure (note that the resistivity for $x = 0.11$ has been divided by 4 to 
  fit in the same scale).
  {\it T$_c$ is defined as the temperature where zero resistance is achieved within experimental accuracy}}
\end{figure}

The optical reflectivity of the samples  was measured approximately every 25~K  from 300~K down to 10~K, 
from 50 to 21000 cm$^{-1}$. The data were collected in a Bruker IFS 66v Fourier transform spectrometer. 
The raw reflectivity data can be found in earlier papers 
\cite{Zimmers05,Santander04}. We show in  fig.2 and fig.3 the optical conductivities of PCCO and   BSCCO 
samples respectively, 
 deduced from the reflectivity spectra through 
a well established fitting procedure \cite{Santander04}.

\begin{figure}
  \begin{center}
      \includegraphics[width=0.9\textwidth]{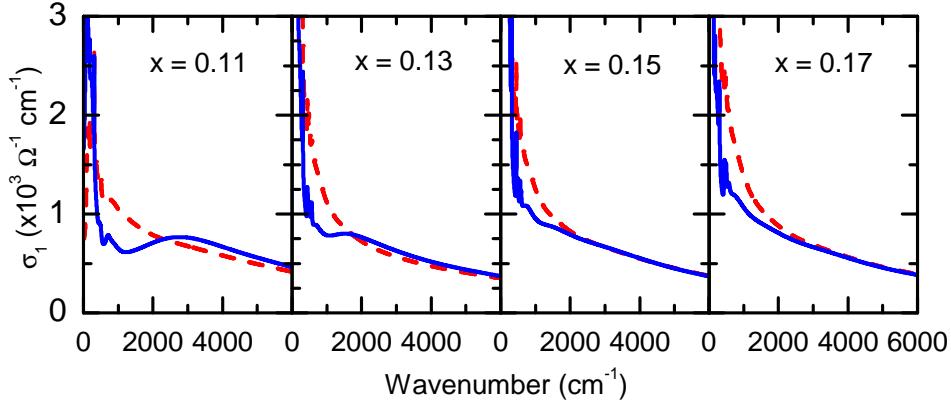}
  \end{center}
  \caption{\label{Fig2}
   Real part of the conductivity of the three PCCO films at 300K (dashed line) and 25K (solid line), 
   {\it i.e.} above $T_c$ for all samples. 
   In the left two panels, a dip-hump structure develops at 25~K. This structure is absent in the right two panels. 
   Note that the scale for $\sigma_{1}(\omega)$ has been expanded in order to make the dip-hump  structure 
   visible;    the low frequency peak centered at zero 
   temperature is therefore heavily cut, but is present in all samples.}
\end{figure}

\begin{figure}
  \begin{center}
      \includegraphics[width=0.9\textwidth]{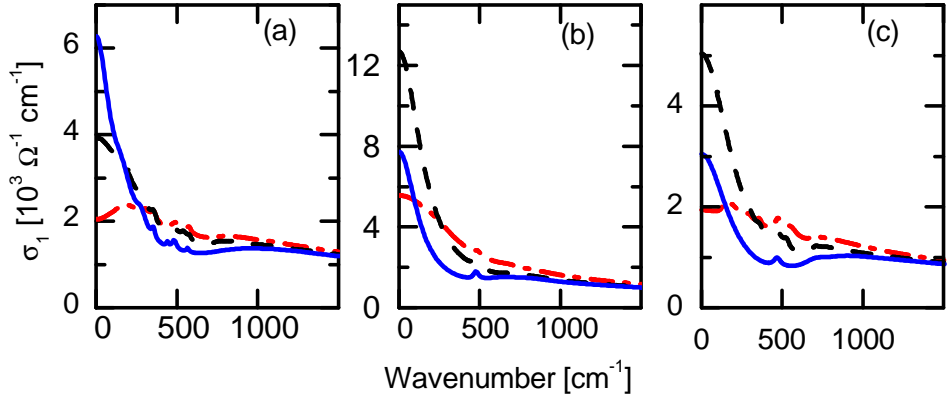}
  \end{center}
  \caption{\label{Fig3}
   Real part of the conductivity of the three BSCCO films.
   (a) underdoped, (b) optimally doped, (c) overdoped. Dash-dotted lines: 250~K (optimally doped 300~K), 
   dashed lines: 100~K (optimally doped 90~K), 
   solid lines : 10~K}
\end{figure}

\section{Spectral weight and normal state partial gap in PCCO and BSCCO}
\label{Spectral weight and NSP gap in PCCO}

All PCCO samples exhibit a low frequency Drude-like peak which narrows with decreasing temperature (fig.2). 
Moving up in frequency, one can see a clear dip-hump structure in the $x=0.11$ and $x=0.13$ 
samples. 
A similar conductivity has been observed in NCCO crystals 
\cite{Onose04,Lupi99}. This dip-hump structure is absent from the 0.17 sample and unclear in the 0.15. 
However, the development of a dip-hump structure by itself is no proof for the opening of a gap: 
the evidence must be looked for in spectral weight transfer.
Indeed, in a conventional metal described by a Drude conductivity, the scattering rate $\tau^{-1}$ 
 decreases with decreasing temperature. Spectral weight being transferred from high to low energies 
as $T$ decreases, the low frequency spectral weight (integrated from $0$ up to a 
frequency $\omega_{2} \simeq \tau^{-1}$) increases at low temperature. 
Conversely, it should decrease at low temperatures in the case of a gap opening 
as a consequence of a low-to-high energy spectral weight transfer. 
In the following discussion, we do not try to estimate the electronic kinetic energy.
We rather use  four different values 
for the upper cut-off frequency $\omega_2$ (and $\omega_{1}=0$, Eq.1) in order to track the 
changes in spectral weight possibly due to the opening of a density gap. 
Figure 4 shows the relative variation of spectral weight defined as 
$\Delta W /W = [W(T) - W(300~K)] / W(300~K)$. 
As expected from the f-sum rule, increasing $\omega_2$ decreases the magnitude of the temperature variation of 
$\Delta W / W$ for all samples. Note however, that for the highest doping, 
$\omega_{2} = 20000 \icm$ is still not large enough to get rid of all temperature dependence. 
Decreasing $\omega_2$ to the 1000--5000 \icm\ range, 
the various samples exhibit very different behaviors. For $x = 0.17$, the spectral weight increases 
steadily as $T$ decreases. This metallic trend is reversed below $\sim $~150~K for $x = 0.13$ and below 
$\sim $~220~K   for $x = 0.11$. This slope inversion is the signature of spectral weight being transferred 
from low to high energy, characteristic of a gap opening. The sample with $x = 0.15$ displays an 
intermediate behavior as the spectral weight levels off at all $\omega_2$ values below $\sim $~50~K, 
thus making the opening  of a gap unclear in this sample.

To shed some light on this latter issue, we plot, in the right panel of fig.4,  $\Delta W / W$  
 with $\omega_2 = 1000 \icm$  for samples $x = 0.13$, 0.15 and 0.17. Cooling from room temperature, 
all samples display an increase of $\Delta W / W$ consistent with a Drude-like peak narrowing. 
In the $x  = 0.13$  film, the gap opening  reverses this trend below $\sim$~150~K, 
 and $\Delta W / W$ decreases. 
 The $x = 0.15$ low frequency spectral weight increases less rapidly than that from $x = 0.17$ 
 below $\sim$~120~K and levels off at low temperature. 
This effect could be attributed to the  thermal evolution of the scattering rate being different in these 
two samples. However, 
as shown in the inset of the right panel from fig.4, the scattering rate of both  $x = 0.15$ and 0.17  samples 
is almost identical and, most importantly, 
the temperature dependence for $x = 0.15$  is featureless. Therefore, the saturation 
observed in $\Delta W /  W$ below $\sim$~120~K strongly suggests that a gap opens below this temperature also 
for $x = 0.15$.

\begin{figure}
  \begin{center}
		\includegraphics[width=0.9\textwidth]{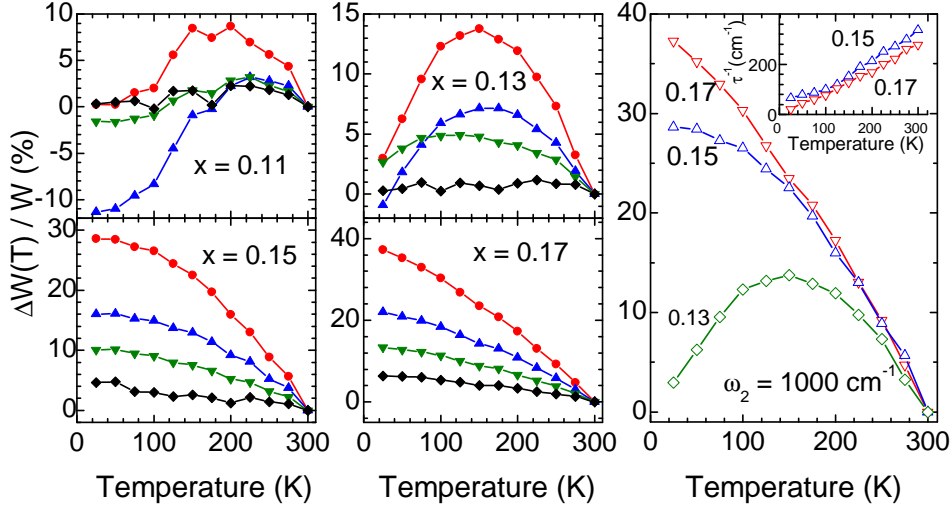}
  \end{center}
  \caption {The four left panels show the relative variation of the spectral weight 
  $\Delta W/W = [W(T)-W(300~K)] / W(300~K)$ for the  PCCO films using $\omega_1 = 0$ and 
  four values for $\omega_2$: 1000 $\icm$ (circles), 2000 $\icm$ (up triangles), 5000 $\icm$ (down triangles) 
  and   20000 $\icm$ (diamonds). The right panel compares the data with 
  $\omega_2 = 1000 \icm$ for samples with $x = 0.13$, 0.15 and 0.17. 
  The inset shows the zero frequency extrapolation of the optical scattering rate $\tau_{-1}$ for the 
  $x = 0.15$ and 0.17 films.}
  \label{fig.4}
\end{figure}
\begin{figure}
  \begin{center}
		\includegraphics[width=0.6\textwidth]{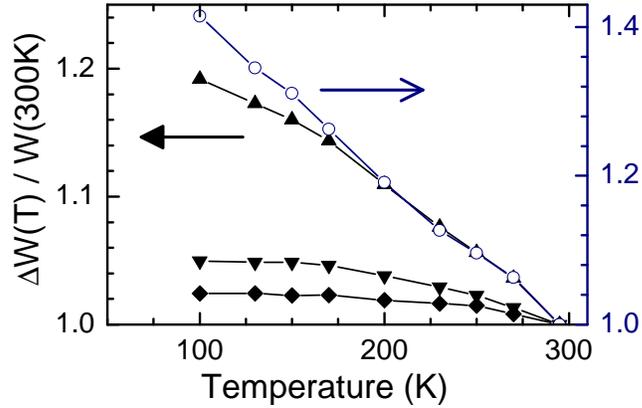}
  \end{center}
  \caption {Relative variation of the spectral weight 
  $\Delta W / W(300 K)$ for the  underdoped BSCCO film using $\omega_1 = 0$ and 
  four values for $\omega_2$: 500~cm$^{-1}$ (open circles), right scale as indicated by the open arrow;  
  1000~cm$^{-1}$ (up triangles), 5000~cm$^{-1}$ (down triangles) and 
  20000~cm$^{-1}$ (diamonds), left scale as indicated by the solid arrow. Note that the temperature scale does 
  not start at zero,   the data points being shown above $T_c$=70~K}
  \label{fig.5}
\end{figure}

The   gap observed  up to $x = 0.15$ is however only partial, {\it i.e.}, it does not open over the whole 
Fermi surface, as fig.2 shows that a Drude-like peak persists at all temperature.
We used the spectral weight plots of fig.4 to define the gap opening temperature and 
to show that this temperature vanishes for a Ce concentration in the range $0.15 < x \leq 0.17$. 
We showed that long range order as seen by neutrons and the optical gap disappear 
at the same Ce concentration. This observation, together with the appearance of a gapped 
Fermi surface \cite{Armitage02}, led us to suggest the occurence of a quantum critical point 
inside the superconducting dome \cite{Zimmers05}.

Independent of the latter interpretation, the opening of a NSP gap in PCCO may be contrasted with its absence 
in BSCCO \cite{Santander02}. We show in fig.5 the temperature dependence of the spectral weight in the underdoped 
BSCCO sample,
where the NSP gap has been observed through many different spectroscopies. In our sample, the onset of this 
NSP gap is located around 150~K, according to the deviation of the resistance from its linear temperature 
dependence (fig.1, left panel). 
We do not show here the temperature dependence of the spectral weight for the other samples: it is 
steadily increasing as the temperature decreases
in a similar way as in PCCO, x=0.17. Therefore, there is no clear-cut sign of the opening of the NSP gap in the 
underdoped p-type material which we have studied. The behavior is similar in underdoped YBa$_2$Cu$_3$O$_{7-\delta}$ (YBCO) 
single crystals (ortho-II YBCO), where the spectral weight also increases as $T$ decreases, up to 
$\omega_2= 15000$~cm$^{-1}$  \cite{Hwang06}.   

  Such different optical properties of n-type and p-type  cuprates are discussed further.

\section{Spectral weight, kinetic energy and its change through $T_c$ as a function of doping in BSCCO}
\label{Spectral weight, kinetic energy and its change through $T_c$ as a function of doping in BSCCO}

Up to this point, we did not make use of the fact that the spectral weight is representative of the electronic 
kinetic energy. We have studied the change of  the spectral weight in three BSCCO samples, that
we have carefully selected and studied \cite{Santander04},  when cooling them from above $T_c$ deep 
down into the SC state. We do not report data for the PCCO samples, because  
we were not able, within our experimental accuracy, to pin down a change of spectral weight between the 
normal and the superconducting state, as discussed further. We explained in section 2 that this spectral 
weight change represents the change of the electronic kinetic energy $E_{kin}$.

We  use $\omega_2$=1~eV and 
$\omega_1 = 0$ or $0^+$, in the normal (N) and superconducting (SC) states respectively (where the superfluid weight is 
 included, {\it i.e} the weight of the $\delta$ function at zero frequency). 
We plot in fig.6 the spectral weight integrated up to 8000~cm$^{-1}$ (1~eV) for the three BSCCO samples. 
 
\begin{figure}
  \begin{center}
      \includegraphics[width=0.9\textwidth]{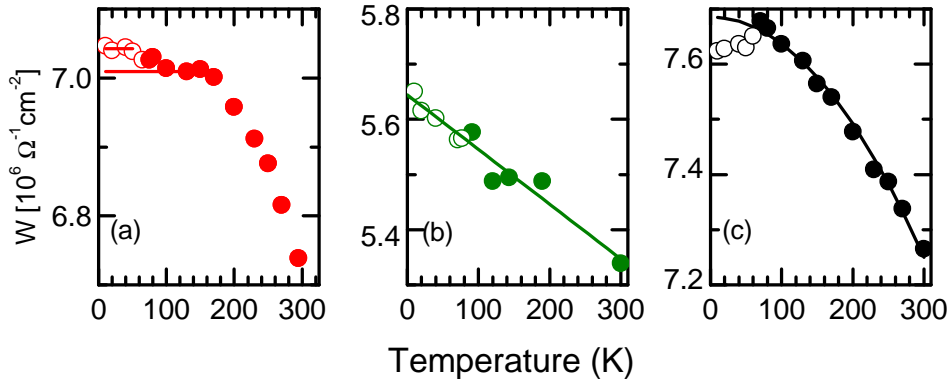}
  \end{center}
  \caption{\label{Fig6}
   Spectral weight integrated up to 1~eV of the three BSCCO films.
   a) underdoped, $T_c$=70~K; b) $\sim$ optimally doped, $T_c$=80~K; c) overdoped, $T_c$=63~K;
   the full symbols are above $T_c$ (integration from $0^+$), the open symbols below $T_c$, 
   (integration from $0$, including the weight of the superfuid). Solid lines are best fits adjusting the 
   temperature dependence in the normal state. a) fit to a constant from 150~K and down; b) fit to a linear 
   dependence from 300~K; c) fit to a $T^2$ dependence. See text.}
\end{figure}

All BSCCO samples exhibit an increase of spectral weight,  {\it i.e} a decrease of kinetic energy down to $T_c$ 
(note that our definition of $T_c$ is where the resistance drops to zero within the experimental uncertainty -- 
see fig.1, left panel). The open symbols in fig.6 refer to temperatures below $T_c$, but the resistance 
starts to drop at a higher temperature (fig.1, left panel). 
In the underdoped (fig. 6-a) and in the overdoped (fig. 6-c) samples, we observe a break  approximately 
at $T_c$ in the slope of the 
spectral weight as a function of temperature.  Remarkably, this break is followed by an increase of 
spectral weight in the SC state in the underdoped case, and a decrease (the sign of the slope is reversed) 
in the overdoped case.  For the optimally doped sample, no clear conclusion can be drawn. 

The extrapolation obtained by adjusting the data points in the N state (full symbols) to a $T^2$ dependence, 
which we used in the overdoped case, is often mentioned and also shown in the literature 
\cite{Molegraaf02,Ortolani05}. We have also tried for this latter sample (not shown) a linear fit (which turns out to be 
worse than the $T^2$ one); 
it is clear that whatever the fit, the normal state extrapolates at a larger value than the one of the SC state:  
there is a net decrease of spectral weight in the SC state with respect to the 
N state, thus an increase of kinetic energy. This is the standard BCS expected behavior, and we made it 
quantitative in an earlier paper \cite{Deutscher05}.
In  the underdoped sample, panel (a), fig.6, 
there is no way to adjust the temperature dependence of the spectral weight in the 
300--100~K range to a simple power law.  
 The spectral weight levels off 
around 150~K, and stays constant. We thus select what looks as the simplest extrapolation, namely a constant 
\cite{Note3}. 
Below $T_c$, there is a net 
increase of spectral weight, hence a decrease of kinetic energy. This is clearly unconventional and is also 
discussed in detail in our 
previous paper \cite{Deutscher05}. In optimally doped BSCCO, panel (b), fig.6, we have  fewer temperature points than in the other samples; 
we  extrapolate with a linear behavior. This linear behavior goes through the experimental points 
down to zero, hence we are unable to conclude for this sample, as noted earlier \cite{Deutscher05}. 

As far as the extrapolation of the normal state temperature dependence is needed to obtain quantitative 
estimates of the
change $\Delta E_{kin}$ of kinetic energy, overall there seems to 
be no universal behavior for this temperature 
dependence: for example, in underdoped orthoII-YBCO, the spectral weight increases 
linearly as the temperature is decreased from 300~K down to 60~K ($T_c$=56~K)\cite{Hwang06}. 
In underdoped BSCCO crystals, it increases as $T^2$ 
\cite{Molegraaf02}, but this is at odds with our finding in the underdoped film.
Actually, the question of the temperature dependence of the normal state spectral weight or kinetic energy 
remains unsettled from a theoretical point of view. It could be  be $T^2$ on general grounds through 
a Sommerfeld expansion.
Such a behavior is argued to hold  for cuprates, but then one has to call for two energy 
scales \cite{Ortolani05}. An alternative suggestion is to relate the 
temperature dependence of the normal state spectral weight to the one of the scattering rate 
\cite{Knigavko04,Schachinger05,Carbottepp}. 
Whatever the answer, the experimental results point clearly to a sign change of $\Delta E_{kin}$
through $T_c, $ when sweeping the phase diagram. 
Whether this sign change  relates to the superconductivity 
mechanism is a matter of discussion. 
The anomalous behavior of the underdoped sample has been related to the high 
value of the scattering rate \cite{Karakozov02}.
Results close to the observed behavior throughout the phase diagram have been derived 
in the context of a gapped scattering rate due to the onset of a sharp peak as the system enters the SC 
state \cite{Norman02}. One possible interpretation of this peak is the magnetic resonance mode 
\cite{Hwang06,Timusk04}. 
A  phenomenological approach 
(a large drop of the scattering rate as the system enters the SC state) yields the unconventional behavior 
of the underdoped sample \cite{Schachinger05,Marsiglio06}. In the overdoped sample, one would not expect 
 such a  behavior in particular due to the absence of the resonant mode \cite{Carbottepp}). A Dynamical Mean Field 
Theory  calculation in the framework of the 2D 
attractive Hubbard model yields trends similar to our findings  \cite{Kyung05}.
Finally, a recent calculation in the frame work of Cellular Dynamical Mean Field theory generates the correct trend
 and orders of magnitude for $\Delta E_{kin}$ through the phase diagram \cite{Haule06}.

\section{Discussion}

In this section, we address two main issues which we believe are relevant to our understanding of cuprates.

The first one is the normal state partial (NSP) gap. Several clear-cut differences between p and n-type cuprates were observed. 
Besides the well established 
fact that the gap or hot spot seen by ARPES is not located at the same k vector in both families, a NSP gap is 
not  seen through optical conductivity as a loss of states -- hence of spectral weight--  in p-type cuprates,  
whereas it is clearly observed in n-type cuprates \cite{Note4}.
We argued earlier that the absence of any spectral weight loss in p-type cuprates could be due to an 
anisotropic Fermi velocity \cite{Santander02}. More generally, 
there seems to be a consensus on the transport (DC and AC) measurements being more sensitive to the nodal 
direction than to the anti-nodal direction \cite{Corson00,Devereaux03}. This remains to be confirmed.
 There is no serious issue of the kind in n-type cuprates since the optical data display a clear loss of 
 spectral weight. Therefore, the open question is to know whether the difference between n and p-type cuprates 
 is related to some details of the Fermi surface, with the underlying physics being the same (density gap, 
 quantum critical point) -- these are very controversial issues -- or whether there is an asymmetry 
 if doping with holes or electrons --
 the physics turning out to be different in the two families. Since in both cases, the parent compound 
 is a Mott-type insulator, Mott physics is present in both problems. Indeed, we argued in an earlier 
 paper that n-type cuprates are as strongly correlated as p-type cuprates \cite{Millis05}. 
Recent theoretical work 
 discusses these differences in relation with density wave correlation \cite{Onufrieva05}.
 It is no news that the whole issue of the NSP gap (or "pseudogap") is presently unsettled.

 The second point  is the issue of the relevant energy scales in cuprates.
 In Fermi liquid theory, the whole physics is described by the low energy degrees of freedom. 
 The only relevant energy scale is the Fermi energy $E_F$ (and in the SC state, the gap $\Delta$). 
 We did not discuss in this paper  yet
the energy scale over which one has to integrate in order to exhaust the so-called 
Ferrel-Glover-Tinkham (FGT) sum rule \cite{FGT58,FGT59}; the FGT sum rule states that the area lost at finite 
frequency in the SC state is recovered in the superfluid condensate, when integrating  (in 
BCS superconductors) up to a few times $\Delta$. Deviations above this energy scale are vanishingly small
 and only related to $\Delta $ \cite{Karakozov02}. 
Figure 6 implies that in the underdoped BSCCO sample, the FGT sum rule 
is violated up to 1~eV  \cite{Santander03,Santander04}. 
A systematic study  of this violation  as a function of the upper integration limit
shows that it
 disappears within the experimental uncertainties above 1.5~eV \cite{Santander06}.
 In other terms, the energy scale of recovery of the spectral weight transferred into the $\delta$--function is
 large, $\sim$ 60~ times the SC gap $\Delta$ \cite{Santander03,Deutscher05}. 
 This comes close to the energy of the  charge-transfer (CT) band (not strictly speaking the upper Hubbard band) 
 whose onset is commonly found at $\sim$~1.5~eV 
 \cite{Orenstein90,Uchida91}. Although the energy of the CT band is not actually known for BSCCO 
 since the insulating 
 parent  compound does not exist, we  assume that it is close to  
 La$_{2-x}$Sr$_x$CaCuO$_{4-\delta}$ (LSCO) and YBa$_2$Cu$_3$O$_{7-\delta}$ (YBCO).
 The very fact that the FGT sum rule calls for an anomalously large scale in order to be 
 fulfilled, at least in the underdoped sample, suggests very strongly the relevance of the
 Mott physics with superconductivity.
 In the overdoped sample, we find that the rate of recovery of the spectral weight occurs at a conventional 
 energy scale; this seems to go together with the change of kinetic energy being also  conventional, both 
 in sign and in magnitude \cite{Santander03,Santander04,Deutscher05}. 
 The decrease of spectral weight persists up to our highest reliable frequency (1.5~eV). 
It is only the experimental uncertainty  which sets the energy scale beyond which the decrease of spectral 
weight can be seen.  We  speculate that in the overdoped regime, Mott physics becomes unimportant.

\section{Conclusion}
\label{Conclusion}

We have discussed in detail how a systematic analysis of spectral weight,  using various 
cut-off frequencies, yields accurate information about a partial gap opening and its change with doping: 
this case was illustrated through our study of PCCO samples, and emphasizes one major difference 
between n-type and p-type cuprates, namely a normal state partial gap being observed in the former and not 
in the latter.
We then took advantage of the identification of the spectral weight to electronic kinetic 
energy  
in the framework of cuprates, if described by a single band model. In case of overdoped BSCCO, the  
identification  of spectral weight and $E_{kin}$ 
works remarkably well, yielding the proper order of magnitude, in 
the framework of a conventional BCS behavior, of the change $\Delta E_{kin}$ of the electronic kinetic energy 
at $T_c$. In the  underdoped sample, 
the  {\it sign} of this change  is reversed, which is unconventional with respect 
to BCS behavior.   
This trend is definitely robust when changing the cut-off frequency. We finally  noted the relevance
of an energy scale close to the one of the charge-transfer band. This is taken as an indication of 
the key role of Mott physics in cuprates. 

\section{Acknowledgements}
\label{Acknowledgements}

We wish to thank H.~Raffy and her co-workers, R.~Greene and his coworkers, for providing us with 
 high quality films. We are grateful to J.~Carbotte, G.~Kotliar, A.~Millis,  M.~Norman, P.~Phillips and
 A-M.~Tremblay  for enlightening discussions.



\begin{thebibliography}{00}
\bibitem{note1} Such regions are also often called 'hot spots' if ARPES intensity drops sharply with 
respect to a clear Fermi surface.
\end{thebibliography}
\end{document}